\documentclass[aps,prd,onecolumn,floats,showpacs, nofootinbib]{revtex4}
\usepackage{graphics,epsfig,color} 
\usepackage{amsmath,amssymb}

\begin{document}
\title{New derivation of the Lagrangian of a perfect fluid with a barotropic equation of state}
\author{Olivier Minazzoli}
\affiliation{Jet Propulsion Laboratory, California Institute of Technology,\\
4800 Oak Grove Drive, Pasadena, CA 91109-0899, USA}
\author{Tiberiu Harko}
\affiliation{Department of Physics and Center for Theoretical and Computational Physics, The University of Hong Kong,\\
 Pok Fu Road, Hong Kong, People's Republic of China}
\begin{abstract}
In this paper we give a simple proof that when the particle number is conserved, the Lagrangian of a barotropic perfect fluid  is $\mathcal{L}_m=-\rho \left[c^2 +\int P(\rho)/\rho^2 d\rho \right]$, where $\rho$ is the \textit{rest mass}  density and $P(\rho)$ is the pressure. To prove this result nor additional fields neither Lagrange multipliers are needed. Besides, the result is applicable to a wide range of theories of gravitation. The only assumptions used in the derivation are:  1) the matter part of the Lagrangian does not depend on the derivatives of the metric, and 2)  the particle number of the fluid is conserved ($\nabla _\sigma (\rho u^\sigma)=0$).
\end{abstract}
\pacs{04.20.Cv; 04.20.Fy; 04.50.Kd}

\maketitle


\section{Introduction}

In order to obtain solvable equations of motion, recently developed alternative theories of gravitation use some specific forms of the perfect fluid Lagrangian \cite{coupl, bertolamiPRD08,sotiriouCQG08,faraoniPRD09, faraPLB12}. Most of these approaches start from the work of Brown \cite{brownCQG93}, where the on-shell perfect fluid Lagrangian $\mathcal{L}_m$, without elastic energy,  is shown to reduce  in General Relativity (GR) to $\mathcal{L}_m=-\rho$ , where $\rho$ is the energy density of the fluid. This result is obtained  by introducing  various additional fields, as well as Lagrange multipliers in order to effectively be able to reconstruct a perfect fluid stress-energy tensor concordant with the laws of thermodynamics -- such as the matter current conservation. However, because of the additional fields, it turns out that the on-shell Lagrangian can also be $\mathcal{L}_m=P$, where $P$ is the pressure of the fluid. This degeneracy of the Lagrangian has no consequences in GR, since both Lagrangians lead to the same equations of motion. In \cite{Taub} was proposed an alternative perfect fluid Lagrangian which is function of the hydrodynamic variables $u^{\alpha }$, $\rho $, and $T$, where $u^{\alpha }$ is the fluid four-velocity, and $T$ the rest temperature of the fluid, and of the gravitational field variables $g_{\mu \nu}$. Also, the equations of hydrodynamics for a perfect fluid in general relativity have been cast in Eulerian form, with the four-velocity being expressed in terms of six velocity potentials in \cite{Schutz1}. The velocity-potential description leads to a variational principle whose Lagrangian density for the perfect fluid is the pressure $P$. Let us also note that a matter Lagrangian of the form $\mathcal{L}_m=-\rho \left(1+\epsilon\right)$, where $\epsilon $ is the elastic potential (or the internal energy) was considered in \cite{Hawking} to derive the equations of motion of the perfect fluid from a variational principle. Otherwise, variational principles for perfect and imperfect general relativistic fluids were considered in \cite{Bailyn}.

However, it turns out that in some alternative theories of gravity, the matter Lagrangian appears explicitly in the field equations. Therefore, the field equations of those theories seem to be different whether one considers $\mathcal{L}_m=-\rho$ or $\mathcal{L}_m=P$. Hence, in these models the Physics seem to be different depending on this choice -- which is not satisfactory with respect to the monistic view of modern Physics, which requires a unique mathematical description of the natural phenomena.

The first obvious thing one can claim about this situation is that there is absolutely no reason why the results obtained in GR by Brown \cite{brownCQG93}  should be applicable in the theories where the matter Lagrangian enters directly in the field equations. Even more than that, the simple fact that the laws of Physics depend on the considered Lagrangian  should be viewed as a proof of the non-applicability of Brown's results in theories where the degeneracy of the matter Lagrangian leads to a variety of different field equations. Therefore, all works \cite{bertolamiPRD08, sotiriouCQG08, faraoniPRD09, faraPLB12,scalar} considering that one can write $\mathcal{L}_m=P$ for their on-shell perfect fluid Lagrangian -- or any linear combination of $-\rho$ and $P$ -- may be incorrect as long as they deal with theories where the matter Lagrangian enters directly in the field equations (unless one can prove the opposite in some specific situation).

To be more specific, two main theories have been considered where $\mathcal{L}_m$ enters directly in the
field equations: $f(R)$ theories with some non-minimal matter/curvature coupling
\cite{coupl, bertolamiPRD08,sotiriouCQG08,faraoniPRD09,harkoPRD10};
or Brans-Dicke-like scalar-tensor theories with some non-minimal matter/scalar coupling
\cite{faraPLB12, scalar}.

In a recent work \cite{harkoPRD10}, instead of using the results of Brown \cite{brownCQG93} and of trying to apply them to the theory under consideration, the Lagrangian for a barotropic perfect fluid was derived from the equations of motion induced by the adopted action. In addition, the conditions of the conservation of the matter fluid current ($\nabla _\sigma (\rho u^\sigma)=0$, where $u^\alpha$ is the 4-velocity of the fluid, and $\rho $ is the rest mass energy density), as well as of the non-dependency of the matter Lagrangian with respect to the derivatives of the metric were also imposed. By using these assumptions one can show that for the considered modified gravity model one has $\mathcal{L}_m=-\rho \left[1 +\int P(\rho)/\rho^2 d\rho \right]$.

In the following, we extend the result of \cite{harkoPRD10} to any gravitational theory that satisfies the conditions of the conservation of the matter fluid current as well as the non-dependency of the matter Lagrangian with respect to the derivatives of the metric. Besides, the present demonstration is actually much simpler than in \cite{harkoPRD10}. Also, in the Appendix we show that in the case of scalar-tensor theories with scalar field/matter coupling the result is compatible with a Brown-like way of deriving the Lagrangian .

In the present paper we use the Misner-Thorne-Wheeler  (MTW) conventions \cite{MTW}. Also, while $c=1$ has been used in the Introduction Section in order to match previous studies' notation, we will explicitly keep $c$ in the rest of the paper.

\section{The Lagrangian of a barotropic perfect fluid}
\label{sec:deriv}
We  start with the usual definition of the stress-energy tensor $T_{\mu \nu}$, given by
\begin{equation}
\label{eq:STo}
T_{\mu \nu}=-\frac{2}{\sqrt{-g}} \frac{\delta(\sqrt{-g}\mathcal{L}_m)}{\delta g^{\mu \nu}}.
\end{equation}

Considering the usual assumption that the matter part of the Lagrangian $\mathcal{L}_m$ does not depend on the derivatives of the metric, we obtain
\begin{equation}
\label{eq:SEt1}
T_{\mu \nu}=-\frac{2}{\sqrt{-g}} \frac{\partial(\sqrt{-g}\mathcal{L}_m)}{\partial g^{\mu \nu}}=\mathcal{L}_m g_{\mu \nu} - 2 \frac{\partial \mathcal{L}_m}{\partial g^{\mu \nu}}.
\end{equation}

Now, by considering a fluid with a barotropic equation of state $P(\rho)$, we can assume that $\mathcal{L}_m$ depends on $\rho$ only. If one considers that the matter current is conserved ($\nabla _\sigma (\rho u^\sigma)=0$), then one can prove that \cite{fockBOOK64,harkoPRD10}:
\begin{equation}
\label{eq:srho}
\delta \rho = \frac{1}{2} \rho \left(g_{\mu \nu} - u_\mu u_\nu  \right) \delta g^{\mu \nu},
\end{equation}
where $u^\alpha$ is the 4-velocity of the fluid, defined in a system of coordinates  $x^\sigma$ as $u^\alpha=dx^\alpha/ds$, where $ds$ is such that $ds^2=-c^2 d\tau^2$, with $\tau$ the proper time of the fluid particles, and with $\rho c^2$ the \textit{rest mass energy density}. Using Eqs.~(\ref{eq:SEt1}) and (\ref{eq:srho})  we obtain \cite{harkoPRD10}
\begin{equation}
T^{\mu \nu}= \rho \frac{d \mathcal{L}_m}{d \rho} u^\mu u^\nu + \left( \mathcal{L}_m - \rho \frac{d \mathcal{L}_m}{d \rho} \right)g^{\mu \nu}.
\end{equation}

Now, since we want to obtain the Lagrangian of a barotropic perfect fluid, we have to equate this equation with the usual stress-energy tensor of a barotropic perfect fluid,
\begin{equation}
T^{\mu \nu}= -\left[\epsilon(\rho)+P(\rho) \right] u^\mu u^\nu + P(\rho)g^{\mu \nu},
\end{equation}
where $\epsilon (\rho )$ is the total energy density of the fluid.
Therefore we obtain the following two equations,
\begin{eqnarray}
\mathcal{L}_m &=& -\epsilon(\rho),\label{eq:solLM}\\
\frac{d \mathcal{L}_m}{d \rho}&=& -\frac{\epsilon(\rho)+P(\rho)}{\rho}. \label{eq:solpLM}
\end{eqnarray}
Using Eqs.~(\ref{eq:solLM}) and (\ref{eq:solpLM}), we obtain the following first order linear differential equation for the energy density of the fluid,
\begin{equation}
\frac{d \epsilon(\rho)}{d \rho}= \frac{\epsilon(\rho)+P(\rho)}{\rho}.
\end{equation}
The general solution of this equation is 
\begin{equation}
\epsilon(\rho)=C \rho + \rho \int \frac{P(\rho)}{\rho^2} d\rho, \label{eq:eps}
\end{equation}
where $C$ is an arbitrary integration constant. Therefore, we have shown that the Lagrangian
\begin{equation}
\label{eq:LagrangianwC}
 \mathcal{L}_m=- C \rho - \rho \int \frac{P(\rho)}{\rho^2} d\rho,
\end{equation}
 leads to an energy-momentum tensor of the form
\begin{equation}
 T^{\mu \nu}=- \left\{\rho \left[C+\Pi(\rho) \right]+P(\rho) \right\} u^\mu u^\nu + P(\rho) g^{\mu \nu},
\end{equation}
where
\begin{equation}
\Pi(\rho)=\int \frac{P(\rho)}{\rho^2} d\rho= \int \frac{dP}{\rho} - \frac{P}{\rho}, \label{eq:Pirho}
 \end{equation}
is  the elastic compression potential energy per unit mass of the fluid \cite{fockBOOK64}. The integration constant is given by $C=c^2$. A simple way to figure it out is to take the point particle limit of the considered action, and to equate it with the usual point particle action ($S_m= mc^2 \int ds$, where $m$ is the rest mass of the point particle). Therefore, the Lagrangian of a barotropic perfect fluid is given by
\begin{equation}
\label{eq:Lagrangian}
 \mathcal{L}_m=- \rho \left( c^2 + \int \frac{P(\rho)}{\rho^2} d\rho \right).
\end{equation}
The corresponding stress-energy tensor can be written as
\begin{equation}
 T^{\mu \nu}= - \left\{\rho \left[c^2+\Pi(\rho) \right]+P(\rho) \right\} u^\mu u^\nu + P(\rho) g^{\mu \nu},
\end{equation}
or, equivalently,
\begin{equation}
 T^{\mu \nu}=  \left\{\rho \left[c^2+\Pi(\rho) \right]+P(\rho) \right\} U^\mu U^\nu + P(\rho) g^{\mu \nu},
\end{equation}
where $U^\alpha=c^{-1} dx^\alpha/d\tau$  is the four-velocity of the fluid divided by the speed of light \footnote{It also has to be pointed out that, conversely to other works, a correct Lagrangian is used in \cite{moiPRD12}.}. One can verify that this stress-energy tensor is indeed of the form assumed, for instance, in celestial relativistic mechanics \cite{kopejkinCM88,brumbergNCB89,klionerPRD00,Soffel:2003cr,kopeikinPR04}.

Also, from the conservation of the \textit{rest mass} density ($\nabla _\sigma (\rho u^\sigma)=0$) and using equations (\ref{eq:eps}) and (\ref{eq:Pirho}), one derives the usual non-conservation equation for the total energy density:
\begin{equation}
\nabla _\sigma (\epsilon u^\sigma)=-P ~\nabla_\sigma(u^\sigma).
\end{equation}

Let us note that, for a fluid satisfying a linear barotropic equation of state $P=\left(\gamma -1\right)\rho c^2$ with $\gamma ={\rm constant}$, one has:
\begin{equation}
\mathcal{L}_m=-\rho c^2\left[1+(\gamma -1)\ln\frac{\rho }{\rho _0}\right],
\end{equation}
where $\rho _0$ is an arbitrary constant of integration. The specific example with the stiff fluid equation of state $\gamma =2$ gives $\mathcal{L}_m=-\rho c^2\left[1+\ln\frac{\rho }{\rho _0}\right]$.

Otherwise, for a fluid satisfying a polytropic equation of state $P=K \rho^{1+1/n}$, where both $K$ and $n$ are constant, one gets:
\begin{equation}
\mathcal{L}_m=-\rho \left[c^2+K \rho^{1/n}+ C \right],
\end{equation}
where $C$ is an integration constant.

\section{Conclusions and final remarks}

In this note, we have shown that as long as one considers (generic) cases where both equations (\ref{eq:SEt1}) and (\ref{eq:srho}) are valid, in the MTW signature the Lagrangian of a barotropic perfect fluid  is
\begin{equation}
 \mathcal{L}_m=- \rho \left[ c^2 + \int \frac{P(\rho)}{\rho^2} d\rho \right],
\end{equation}
regardless of the nature of the theory considered otherwise.

However, it has to be emphasized that even if the conservation of the matter  current is a legitimate assumption, a non-conservation is conceivable as well. Indeed, the theories considered for our purpose all lead to a non-conservation of the stress-energy tensor. In the case of the scalar-field theories with scalar field/matter coupling for instance, it means that there is an energy transfer between the scalar and the matter fields. The energy transfer has two implications: either it modifies the geodesic equation of motion of the free particles, or it induces a non-conservation of the particle number (or both at the same time). Therefore, it would be interesting to relax the condition on the conservation of the matter current in order to obtain a more general result.

Otherwise, in this paper we prove that the works \cite{bertolamiPRD08, sotiriouCQG08, faraoniPRD09,faraPLB12, scalar} that use $\mathcal{L}_m=P$, or any of the specific linear combination of $-\rho$ and $P$ for the Lagrangian, are incompatible with the matter current conservation. Also, let us remark that it seems very unlikely that in the most general case -- where the matter current conservation constraint is relaxed -- the Lagrangian would reduce precisely to either $\mathcal{L}_m=P$ or any of the specific linear combination of $-\rho$ and $P$ that is used in many works. Therefore, we finally argue that one should be very cautious before considering any of the results of the works that used the results of Brown \cite{brownCQG93} in alternative theories of gravity without questioning the validity of its extrapolation to this type of models,  as, for instance, in \cite{bertolamiPRD08, sotiriouCQG08,faraoniPRD09,faraPLB12, scalar}.

\begin{acknowledgments}
This research was supported by an appointment to the NASA Postdoctoral Program at the Jet Propulsion Laboratory, California Institute of Technology, administered by Oak Ridge Associated Universities through a contract with NASA. \copyright 2012 California Institute of Technology. Government sponsorship acknowledged.
\end{acknowledgments}


\appendix

\section{On the extension of the work of Brown to scalar-tensor theories with scalar/matter coupling}
The action describing Brans-Dicke theory with a universal spin-0/matter coupling can be written as follows \cite{moiPRD12},
\begin{equation}\label{eq:action}
S=\int d^4x\sqrt{-g} \left( \Phi R -
\frac{\omega}{\Phi} \left(\partial_{\sigma}\Phi \right)^2 -
V(\Phi) +2 f(\Phi)\mathcal{L}_m \right),
\end{equation}
where $g$ is the metric determinant, $R$ is the Ricci scalar
constructed from the metric $g_{\mu \nu}$, and $\mathcal{L}_m$ the matter Lagrangian. One possible Lagrangian that describes a perfect fluid without elastic energy in GR is given by \cite{brownCQG93}:
\begin{eqnarray}
\mathcal{L}_m &=&-\epsilon\left(|J|/\sqrt{-g},s \right)+\nonumber\\
&&\frac{J^\sigma}{\sqrt{-g}}\left[\partial_\sigma \psi + s \partial_\sigma \theta+ \beta_A \partial_\sigma \alpha^A\right],
\end{eqnarray}
where $J^\alpha$ is the particle flux ($|J|\equiv \sqrt{-g_{\rho \sigma} J^\rho J^\sigma}$), $s$ is the entropy per particle, $\alpha^A$ is the 3 Lagrangian coordinates, and $\psi$, $\theta$ and $\beta_A$ are 6 spacetime scalars. Now, if one directly inserts this Lagrangian into Eq.~(\ref{eq:action}), one obtains the following density,
\begin{eqnarray}
\label{eq:LM1}
L_m &=&\sqrt{-g} f(\Phi) \big\{-\epsilon\left(|J|/\sqrt{-g},s \right)\\
&&+(-g)^{-1/2} J^\sigma\left[\partial_\sigma \psi + s \partial_\sigma \theta+ \beta_A \partial_\sigma \alpha^A\right]\big\}. \nonumber
\end{eqnarray}
However, it is easy to figure out that this density does not lead to the required thermodynamic constraints, such as the particle number conservation, ($\partial_\sigma J^\sigma=0$) and the absence of entropy exchange between neighboring flow line ($\partial_\sigma(s J^\sigma)=0$). Hence, the Lagrangian density (\ref{eq:LM1}) is not suitable to describe a perfect fluid in the class of models considered in this paper. On the contrary, the following Lagrangian density, not only leads to the perfect fluid stress-energy tensor, but also does satisfy the two previously mentioned thermodynamic constraints,
\begin{eqnarray}
\label{eq:LM3}
L_m &=&-\sqrt{-g} f(\Phi) \epsilon\left(|J|/\sqrt{-g},s \right)\\
&&+J^\sigma\left[\partial_\sigma \psi + s \partial_\sigma \theta+ \beta_A \partial_\sigma \alpha^A\right]. \nonumber
\end{eqnarray}
Therefore, it seems that such a Lagrangian density is valid in order to model a perfect fluid Lagrangian -- as long as one wants to impose the matter current conservation $\nabla _\sigma (\rho u^\sigma)=0$ as in the main part of this paper. But then one has to notice that the second term in the right hand side of equation (\ref{eq:LM3}) does not couple neither with the metric, nor with the scalar field. Hence, only the first term in the right hand side of equation (\ref{eq:LM3}) will enter in the field equations. Therefore, the on-shell perfect fluid is simply $\mathcal{L}_m=- \epsilon$, which is in concordance with the main result of this paper.


\begin{thebibliography}{99}

\bibitem{coupl} O. Bertolami, C. G. Boehmer, T. Harko, and F. S.N. Lobo, Phys. Rev. D {\bf 75}, 104016 (2007).

\bibitem{bertolamiPRD08} O. Bertolami, F. S. N. Lobo, and J. Paramos, Phys. Rev.
D {\bf 78}, 064036 (2008).

\bibitem{sotiriouCQG08} T. P. Sotiriou and V. Faraoni, Classical and Quantum
Gravity {\bf 25}, 205002 (2008).

\bibitem{faraoniPRD09} V. Faraoni, Phys. Rev. D 80, 124040 (2009).
\bibitem{faraPLB12} H. Farajollahi, A. Ravanpak, and G. F. Fadakar, Phys.
Lett. B {\bf 711}, 225 (2012).
\bibitem{brownCQG93} J. D. Brown, Classical and Quantum Gravity {\bf 10}, 1579
(1993).

\bibitem{Taub} A. H. Taub, Phys. Rev. {\bf 94}, 1468 (1954).

\bibitem{Schutz1} B. F. Schutz, Phys. Rev. D {\bf 2},  2762 (1970).

\bibitem{Hawking} S. W. Hawking and G. F. R. Ellis, {\it The large scale structure of space-time}, London, University Press, (1973).

\bibitem{Bailyn} M. Bailyn, Phys. Rev. D {\bf 22}, 267 (1980).

\bibitem{scalar} K. Saaidi, A. Mohammadi, and H. Sheikhahmadi, Phys. Rev. D {\bf 83}, 104019 (2011); H. Farajollahi and A. Salehi, Phys. Rev. D {\bf 83}, 124042
(2011); H. Farajollahi and A. Salehi, Journal of Cosmology and
Astroparticle Physics {\bf 7}, 36 (2011).

\bibitem{harkoPRD10} T. Harko, Phys. Rev. D {\bf 81}, 044021 (2010).


\bibitem{MTW} C. W. Misner, K. S. Thorne, and J. A. Wheeler, {\it Gravitation}, San Francisco, W. H. Freeman and Co. (1973).

\bibitem{fockBOOK64} V. A. Fock, {\it The theory of space, time, and gravitation}, New York, Pergamon Press (1959).

\bibitem{moiPRD12} O. Minazzoli,  submitted for publication (2012) (eprint arXiv:1208.2372).

\bibitem{kopejkinCM88} S. M. Kopejkin, Celestial Mechanics {\bf 44}, 87 (1988).

\bibitem{brumbergNCB89} V. A. Brumberg and S. M. Kopejkin, Nuovo Cimento B
{\bf 103}, 63 (1989).

\bibitem{klionerPRD00} S. A. Klioner and M. H. Soffel, Phys. Rev. D {\bf 62}, 024019 (2000).

\bibitem{Soffel:2003cr} M. Soffel, S. A. Klioner, G. Petit, P. Wolf, S. M.
Kopeikin, P. Bretagnon, V. A. Brumberg, N. Capitaine,
T. Damour, T. Fukushima, et al., Astrophys. J. {\bf 126}, 2687 (2003).

\bibitem{kopeikinPR04} S. Kopeikin and I. Vlasov, Physics Reports {\bf 400}, 209
(2004).



\end{thebibliography}
\end{document}